\documentstyle[times,pramana,epsf,floats,amssymb]{ias}

\begin{document}
\mark{{Role of scaling in the statistical modeling of finance}{A.L. Stella and F. Baldovin}}
\title{Role of scaling in the statistical modeling of
  finance\footnote{Based on the Key Note lecture by A.L. Stella at the
    Conference on ``Statistical Physics Approaches to
    Multi-Disciplinary Problems'', IIT Guwahati, India, 7-13 January 2008.}}

\author{Attilio L. Stella and Fulvio Baldovin}
\address{
Dipartimento di Fisica and Sezione INFN, 
Universit\`a di Padova, \\
Via Marzolo 8, I-35131 Padova, Italy.
}


\abstract
{Modeling the evolution of a financial index
as a stochastic process is a problem awaiting a 
full, satisfactory solution since it was first formulated 
by Bachelier in 1900. Here it is shown that the scaling
with time of the return probability density function 
sampled from the historical series suggests 
a successful model. The resulting stochastic process
is a heteroskedastic, non-Markovian martingale, which 
can be used to simulate index evolution on the basis of
an auto-regressive strategy. Results are fully consistent
with volatility clustering and 
with the multi-scaling properties of the return distribution. 
The idea of basing the process construction on scaling, 
and the construction itself, are
closely inspired by the probabilistic renormalization
group approach of statistical mechanics and by a recent
formulation of the central limit theorem for sums
of strongly correlated random variables.
}

\maketitle

\section{Introduction}
Economics and mathematical finance are multi-disciplinary fields
in which the tendency of statistical physicists to focus on
universal laws has been criticized sometimes \cite{ormerod_1}. In
particular, the emphasis on scaling properties typical of many
recent contributions in econophysics has been regarded
with skepticism by some economists, in view of the apparent
scarcity of useful practical consequences of this symmetry \cite{le_baron_1}.

As statistical physicists aware of the key role played by
scaling and universality in the development of the theory of complex
systems in the last decades, we do not share this 
point of view. Universal laws are necessary for building up our
scientific understanding and we do not intend to give them up.
In the specific case of scaling symmetries, it is perhaps fair to
admit that, so far, their potential consequences in finance have
not been fully explored and elucidated \cite{stanley_1}. In the present note we
report on recent work \cite{baldovin_1} demonstrating that scaling, combined with
symmetries enforced by the efficiency of the market, allows  
substantial progress towards the solution of the central problem  
of mathematical finance: assuming that the time evolution
of a financial index, or asset price, amounts to a stochastic
process, formulate a satisfactory model of this process, consistent 
with as many as possible stylized facts established by the
statistical analysis of the historical series 
\cite{cont_1,bouchaud_1,mantegna_1,yamasaki_1,lux_2}. 
This problem awaits
a full, satisfactory solution since it was first formulated by
Bachelier \cite{bachelier_1}.   

The approach sketched below points out far reaching consequences 
of the scaling in time obeyed by the return probability density
function sampled from the historical series of an index. These 
consequences add further strong constraints to those already 
implied by market efficiency, and suggest very plausible
probabilistic rules for the process of index evolution. 
Our goal here is reached through ideas which are partly inspired 
by the probabilistic formulation of the renormalization group
(RG) in statistical mechanics \cite{lasinio_1,kadanoff_1}, and by a recent extension of the
central limit theorem to sums of strongly correlated variables 
obeying anomalous scaling \cite{baldovin_2}. 
To our knowledge, renormalization
group ideas do not seem to have been applied in mathematical
finance so far.

This report is organized as follows. In the second section we
recall the basic facts emerging from the statistical analysis of 
the historical series, taking as example the Dow Jones Industrial
(DJI) index. In parallel we 
also present the core of our derivations and stress 
their links with renormalization group ideas. In the third section
we briefly describe our stochastic model for index evolution, while in the
subsequent, fourth one we review the results of the simulation
of the DJI index. The last, fifth section is
devoted to concluding remarks.
   
\section{Stylized facts and consequences of scaling}

\begin{figure}[htbp]
\epsfxsize=8cm
\centerline{\epsfbox{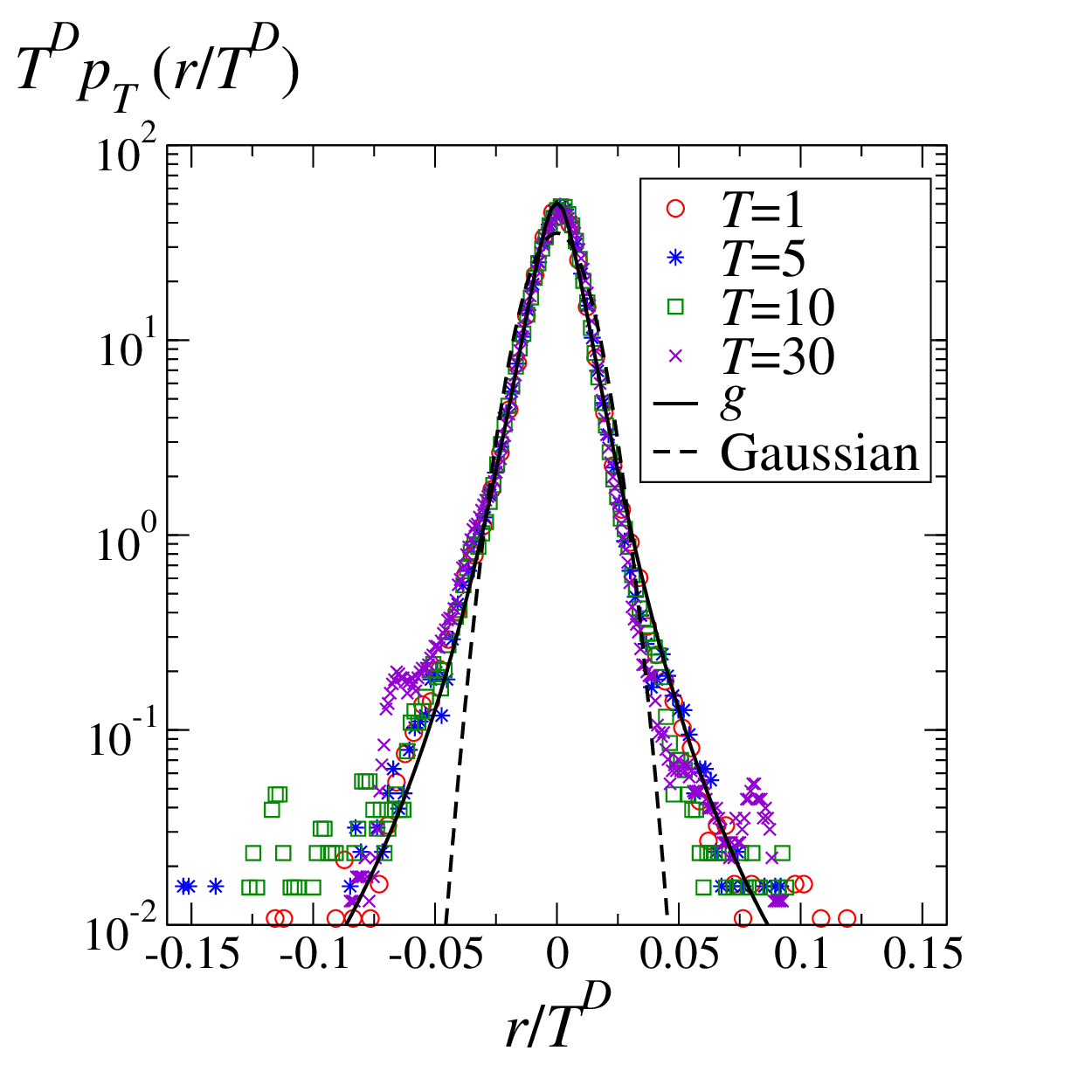}}
\caption{Scaling collapse of the histograms of return PDF's of the DJI. Data
  are sampled over about $26000$ daily closures from 1900 to 2005. The
  dotted curve is a Gaussian, while the continuous one is the best fit
  \protect\cite{sokolov_1} used here to implement our stochastic model.}
\label{fig_scaling}
\end{figure}

Let us indicate by $S(t)$ the value of an index at time $t$. For our 
purposes here, we can assume that $t$ is measured in days and
$S(t)$ represents the daily closure value. A quantity of interest 
\cite{cont_1,bouchaud_1,mantegna_1}
is the logarithmic return $r_{t,T} = \ln(S(t+T)-\ln S(t))$ in the 
interval $[t,t+T]$. 
The values of $\ln S$ used to compute $r_{t,T}$ are assumed to be
detrended, $\ln S(t)\mapsto \ln S(t)-\rho\;t$, where $\rho$ is the
average linear growth over the whole time series. 
A probability density function (PDF) for this
return can be sampled from sufficiently long historical time series. 
The resulting PDF, $p_T(r)$, does depend only on $T$. Indeed, being
sampled with a sliding interval method, $p_T$ conveys only a stationarized
information on return occurrences. For $T$'s in the range from one day
to few months, $p_T$ satisfies approximately simple-scaling:
$p_T(r) = \frac{1}{T^D} g(\frac{r}{T^D})$, where the exponent $D$
turns out be very close to $1/2$ for the indexes of well developed 
markets \cite{di_matteo_1}. 
The scaling function $g$, however, is not Gaussian, and shows power law,
Pareto tails at large $|r|$ \cite{cont_1}. The scaling of $p_T$ in the case of
the DJI index is illustrated by the collapse plot in Fig. \ref{fig_scaling}.
The non Gaussian form of $g$ indicates that successive returns in
the sampling must have strong correlations on the time range
where scaling holds. Before mentioning other stylized facts, it is 
worth concentrating on the scaling of $p_T$, which plays a central role 
in our approach.

In the modern theory of critical phenomena \cite{kadanoff_1}, 
one can consider a finite block 
of $N$ interacting spins and try to identify the critical conditions 
under which doubling the block size ($N \to 2N$) signals the presence of 
scale invariance in the system. In a phenomenological \cite{nightingale_1} version of 
the probabilistic RG approach \cite{lasinio_1}, this
system doubling can be more simply implemented for hierarchical models
in which the Hamiltonian depends only on the total sum, $M$, of the
spins. Here the spins, and thus $M$, are assumed to take values in
$\mathbb R$.
The PDF of $M$ in a system of $N$ spins is then indicated by
$q_N(M)$ and the RG transformation yields $q_{2N}(M)$ as a functional
of $q_N(M)$, once a suitable interaction between blocks is
assumed:
\begin{equation}
q_{2N}(M)=\frac{\int d M_1 d M_2\; e^{H_I(M_1,M_2)}\;q_N(M_1)\;q_N(M_2)
\;\delta(M-M_1-M_2)}
{\int d M_1 d M_2\;e^{H_I(M_1,M_2)}\;q_N(M_1)\;q_N(M_2)}.
\label{prob_rg}
\end{equation}
In Eq. (\ref{prob_rg}),  $H_I( M_1,M_2)$ is the reduced 
(divided by $-k_B T$) coupling between the 
magnetizations of the two blocks. The factor multiplying the delta 
function in the integrand is just the Boltzmann-Gibbs expression
of the joint PDF for the magnetizations $M_1$ and $M_2$ of the
two blocks. Fixed-point critical scaling prevails when the interactions are 
chosen in such a way that Eq. (\ref{prob_rg}) is satisfied by a $q_N$ assuming a
simple-scaling form $q_{2N}(M)=\frac{1}{2^D} q_N(\frac{M}{2^D})$, where 
$D$ is now related to the critical exponents of the model. 

We can envisage a sort of reverse RG strategy in finance. In analogy 
with the magnetic case, the scaling of $p_T(r)$ can be regarded
as a fixed-point scaling for a `block' of $T$ daily returns.
So, we can ask what kind of `coupling' must exist between the returns of
two successive blocks of duration $T$, in such a way that, as we know, $p_{2T}(r)$
satisfies simple-scaling, i.e. $p_{2T}(r)=\frac{1}{2^D} p_T(\frac{r}{2^D})$.
Since there is no Hamiltonian now,
the statistical information on this coupling is embodied in the unknown joint 
PDF of the returns $r_1$ and $r_2$ in the successive intervals, $p_{2T}^{(2)}(r_1,r_2)$. 
Indeed, since the returns $r_1$ and $r_2$ sum up to the return in the interval
of duration $2T$, this joint PDF satisfies:
\begin{equation}
p_{2T}(r)= \int d r_1 d r_2\;p_{2T}^{(2)}(r_1,r_2)\;\delta(r-r_1-r_2).
\label{prob_emp}
\end{equation}
One realizes that $r_1$ and $r_2$ in Eq. (\ref{prob_emp}) play a role analogous to that
of $M_1$ and $M_2$ in Eq. (\ref{prob_rg}). Since $p_{2T}^{(2)}$ is not known,
one can imagine to determine this function in terms of $p_T$ on the basis of Eq. (\ref{prob_emp}).
In the magnetic RG analogy this would amount to determine the critical 
interaction conditions, once the fixed-point scaling form of $q_N(M)$
is given. 
Of course, this determination
is not expected to be unique in general: there can be many different 
$p_{2T}^{(2)}$'s satisfying Eq. (\ref{prob_emp}) for a given $p_T$. However, there are other 
constraints and symmetries helping in the search of the right solution. 
A well established empirical fact is that the average 
\mbox{$\langle r_1 r_2\rangle_{p_{2T}^{(2)}}\equiv
\int d r_1 d r_2 \;p_{2T}^{(2)}(r_1,r_2)\;r_1r_2$} 
must be equal to zero. This can be easily verified
and is in fact an obvious requisite for an efficient market. A deviation from 
zero of this average would open an arbitrage opportunity which would be 
immediately exploited and suppressed by the market. Other constraints
concern the marginal PDF's:
\begin{eqnarray}
p_{T}(r_1)&=& \int d r_2\;p_{2T}^{(2)}(r_1,r_2),
\label{prob_mar_a}\\
p_{T}(r_2)&=& \int d r_1\;p_{2T}^{(2)}(r_1,r_2).
\label{prob_mar_b}
\end{eqnarray}
The validity of Eqs. (\ref{prob_mar_a}-\ref{prob_mar_b}) 
is based on the fact that both $p_T$ and $p_{2T}^{(2)}$
are sampled with a sliding interval method from the historical time series.
This marks a difference with respect to the magnetic case, because there
the marginal PDF's would be computed at a rescaled $N$, due to a 
renormalization effect which is excluded here for the empirical PDF's
of finance. 
On the basis of Eqs. (\ref{prob_emp}-\ref{prob_mar_b}) 
and of the linear decorrelation
of successive returns, it is immediate to derive that, for
the scaling of $p_T$ to occur, one must necessarily have $D=1/2$. It is sufficient to
express the average $\langle(r_1 + r_2)^2\rangle_{p_{2T}^{(2)}}$, and to
take into account that the scaling form of $p_T$ implies that its
second moment must scale as $\langle r^2\rangle_{p_T} \sim T^{2D}$. One then finds immediately
that $2 T^{2D} = (2T)^{2D}$, i.e. $D=1/2$. This result explains the
robustness of the estimate $D \simeq 1/2$ emerging from
the statistical analysis of all indexes in mature markets \cite{di_matteo_1}.

\begin{figure}[htbp]
\epsfxsize=8cm
\centerline{\epsfbox{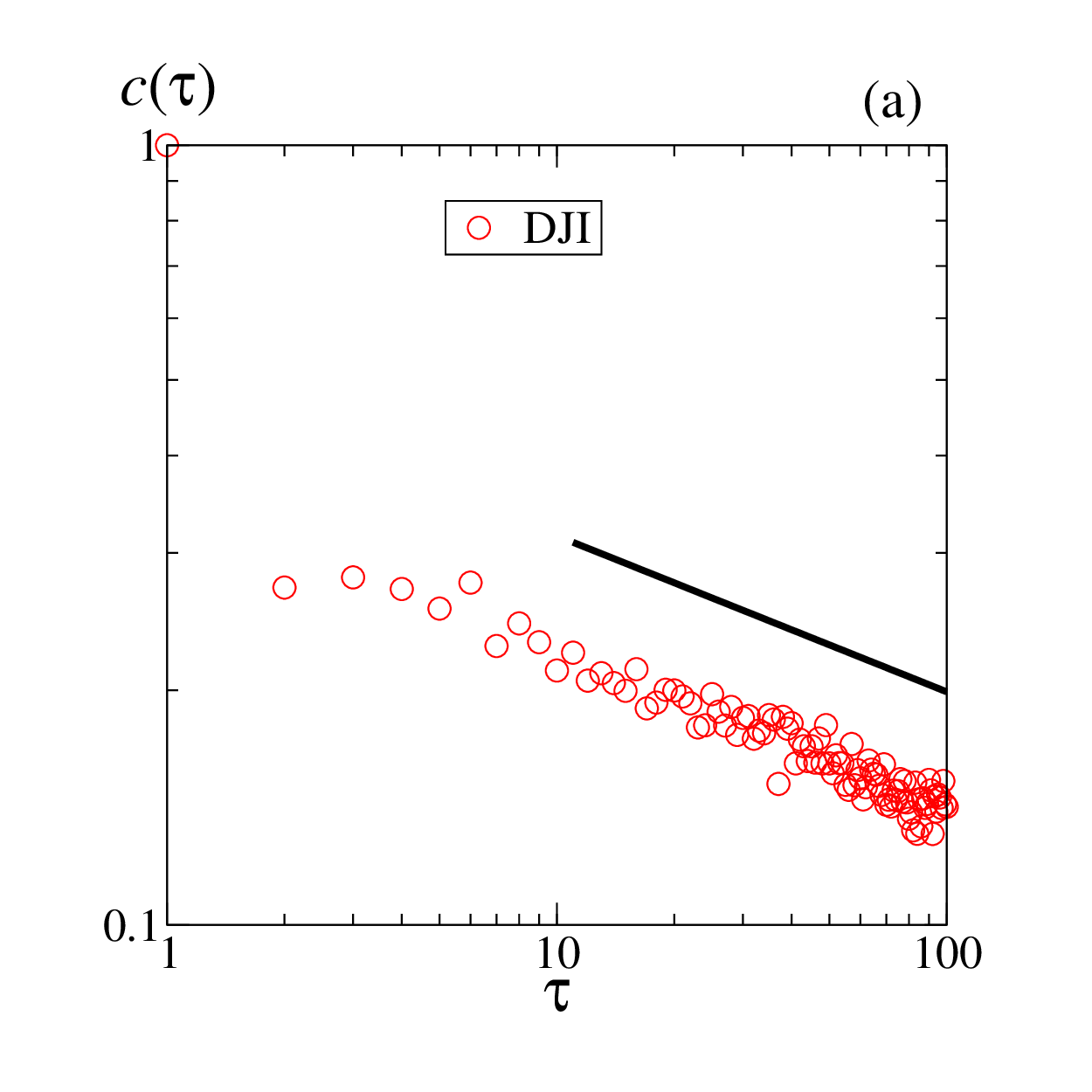}}
\epsfxsize=8cm
\centerline{\epsfbox{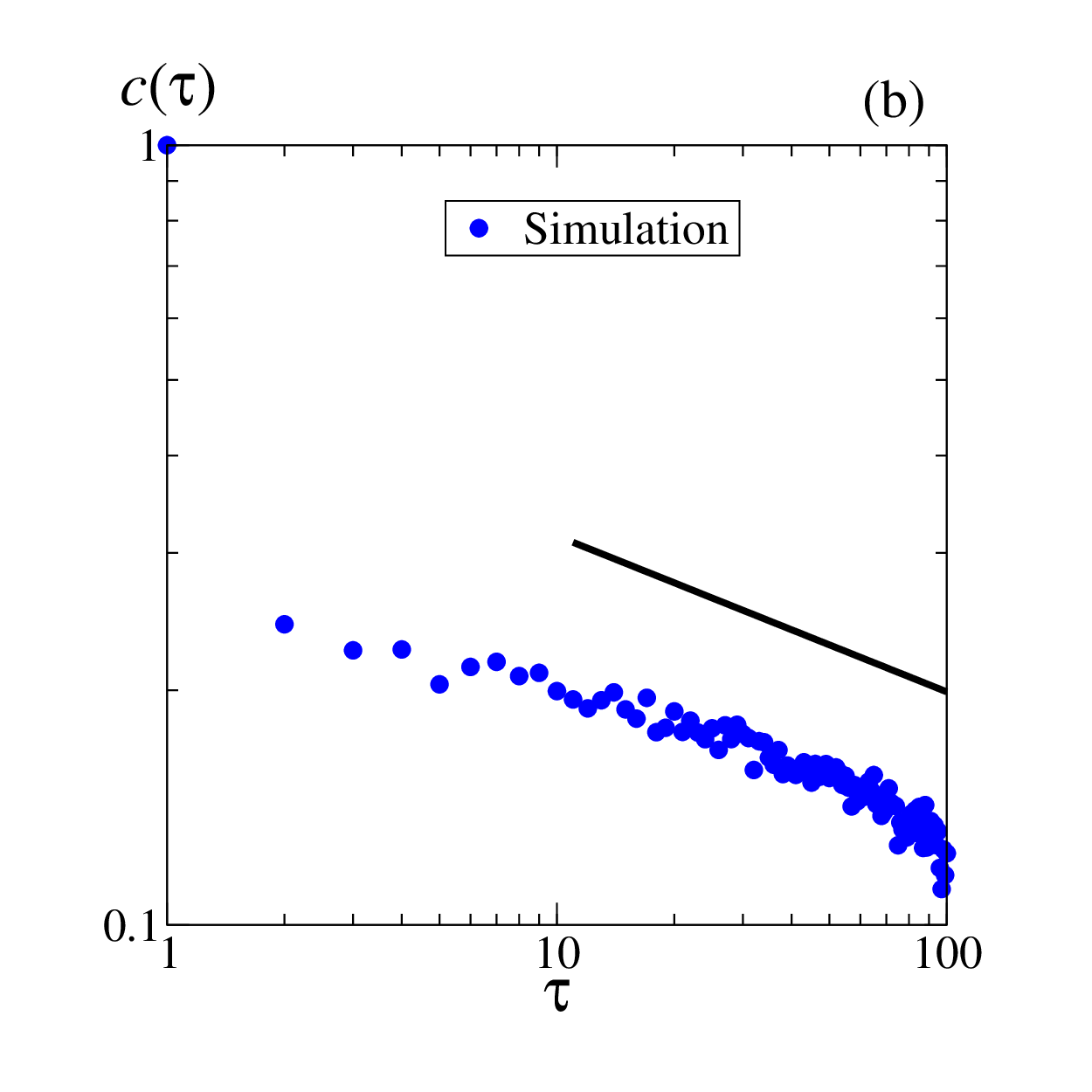}}
\caption{
  (a) Log-log plot of the empirical volatility autocorrelation at time separation $\tau$ (in days),
  \mbox{$
    c(\tau)\equiv\frac{\sum_{t=0}^{t_{max}} |r(t,1)| |r(t+\tau,1)| - 
      \sum_{t=0}^{t_{max}} |r(t,1)|\sum_{t=0}^{t_{max}} |r(t+\tau,1)|/t_{max}}
    {\sum_{t=0}^{t_{max}} |r(t,1)|^2 - [\sum_{t=0}^{t_{max}} |r(t,1)|]^2/t_{max}},
    \label{eq_correlation}
  $}
  where $t_{max}+\tau-1$ is the total length of the time series.
  The data refer to the DJI index and $t_{max}+\tau-1$ is about
  $26000$ days. The continuous line has slope $\simeq -0.2$, the
  exponent of the power-law decay. In (b) the empirical data have been
  replaced by data from the simulation of one history.
}
\label{fig_corr}
\end{figure}

Now let us come back to the problem of expressing $p_{2T}^{(2)}$ in
terms of $p_T$. If the linear decorrelation of successive returns,
i.e. $\langle r_1 r_2\rangle_{p_{2T}^{(2)}} =0$, would imply a complete decorrelation
of $r_1$ and $r_2$, the problem would be easily solved. Indeed, independence
would mean $p_{2T}^{(2)}=p_T(r_1) p_T(r_2)$. By substituting in Eq. (\ref{prob_emp}), and
using the scaling form of $p_T$, we would then conclude immediately that
$D=1/2$ and a Gaussian $g(x)=\exp(-x^2/2\sigma^2)/\sqrt{2\pi\sigma^2}$ 
are necessary for consistency. Indeed, in this case of independence
Eq. (\ref{prob_emp}) 
just imposes to $g$ 
the property of stability which is at the basis of the central limit theorem 
and is satisfied, for finite variance $\sigma^2$, 
by the Gaussian PDF alone \cite{gnedenko_1}. This is even more
directly verified in terms of characteristic functions (CF). For $p_T$
the CF is 
$\tilde p_T(k)=\int d r e^{i k r} p_T(r)= \tilde{g}(T^D k)$, and Eq.(\ref{prob_emp}),
together with the scaling and the independence conditions,
simply reads
\begin{equation}
\tilde{g}((2T)^D k)= \tilde{g}(T^D k)\;\tilde{g}(T^D k),
\label{eq_mult}
\end{equation}
which has 
$\tilde{g}=\exp(-\sigma^2k^2/2)$ as solution for $D=1/2$. We know, however, that the
linear decorrelation of returns is not implying independence:
for example, the so called effect of volatility clustering
leads to $\langle r_1^2 r_2^2\rangle \neq \langle r_1^2\rangle \langle r_2^2\rangle$ for our two successive 
returns. Likewise, a well established fact is that the absolute value of daily returns shows
a strong positive autocorrelation function also at a distance of
months \cite{cont_1,bouchaud_1,mantegna_1}. This autocorrelation function decays
as a power of the time interval $\tau$ 
separating the two days (see Fig. \ref{fig_corr}a).

\begin{figure}[htbp]
\epsfxsize=8cm
\centerline{\epsfbox{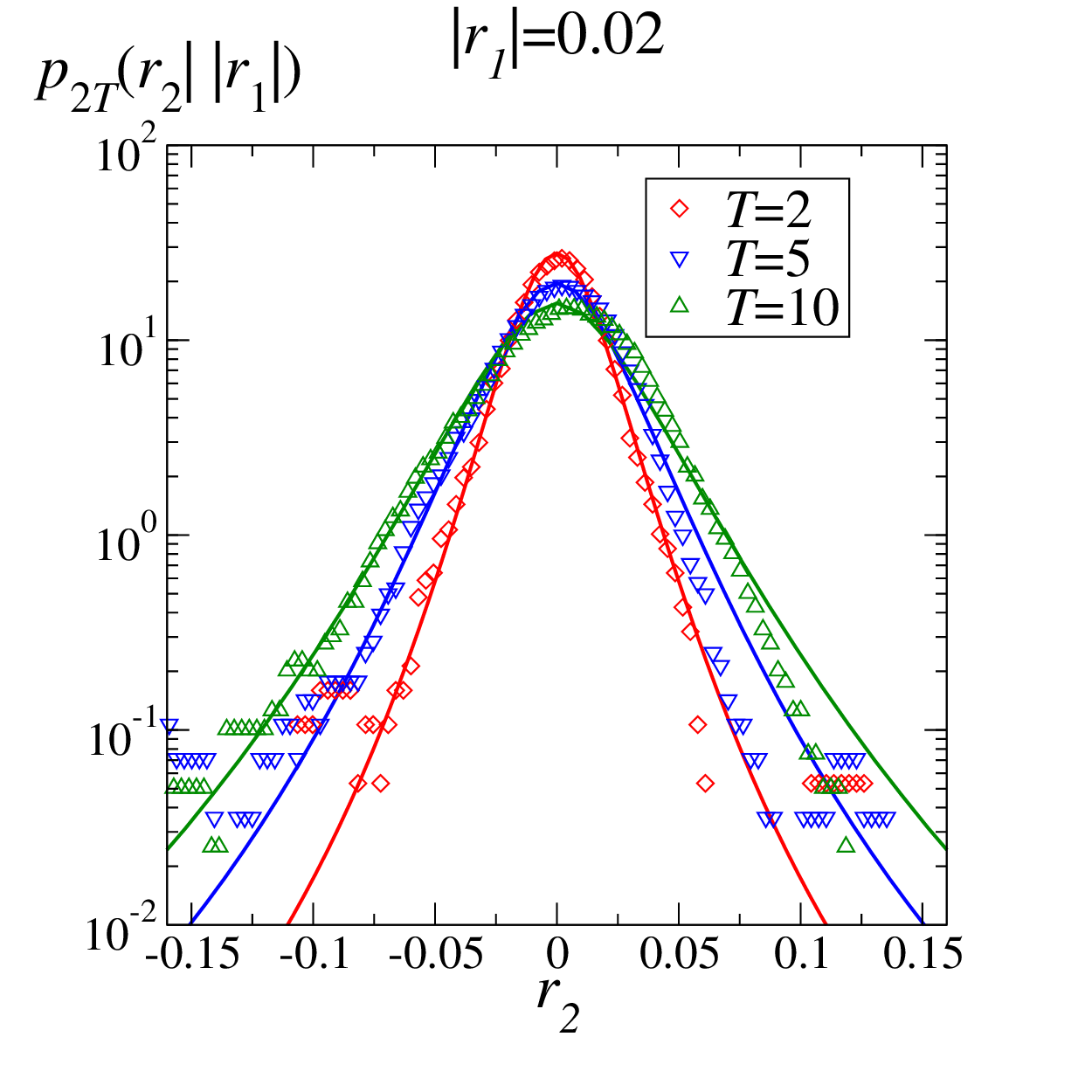}}
\caption{
  Symbols are the empirical conditional PDF, $p_{2T}(r2||r_1|)$, of a return $r_2$ following
  $r_1$ with $|r_1|=0.02$. The continuous curves are obtained on the
  basis of Eq. (\ref{solution_D0p5}), when $\tilde g$ results from the
  fit in Fig. \ref{fig_scaling}
}
\label{fig_conditional}
\end{figure}

Indicating by $\tilde p_{2T}^{(2)}(k_1,k_2)$  the CF
of the joint PDF $p_{2T}^{(2)}$, Eqs. (2-4) above read:
\begin{eqnarray}
\tilde p_{2T}^{(2)}(k,k)&=&\tilde g(\sqrt{2T}\;k),\\
\tilde p_{2T}^{(2)}(k,0)&=&\tilde g(\sqrt{T}\;k),\\
\tilde p_{2T}^{(2)}(0,k)&=&\tilde g(\sqrt{T}\;k),
\end{eqnarray}
where $D=1/2$ has been already substituted. It is then immediate to
realize that a possible solution is simply:
\begin{equation}
\tilde p_{2T}^{(2)}(k_1,k_2)=\tilde g\left(\sqrt{T k_1^2+T k_2^2}\right),
\label{solution_D0p5}
\end{equation}
provided such a $\tilde p_{2T}^{(2)}$ is a characteristic function,
i.e. its inverse-transform is a PDF in $(r_1,r_2)$.
This is not of course the case for any
$\tilde{g}$, but one can show that there is a large class of
CF's satisfying this requisite, as can be checked by
numerics \cite{baldovin_1}, or established on the basis of rigorous
theorems \cite{baldovin_3}. 
The solution in Eq. (\ref{solution_D0p5}) is of course not the unique possibility.
However, it is strongly suggested, in first place, by the
good consistency it demonstrates with the statistical data.
When sampled over the whole history of the DJI index from
$1900$ to $2005$, the histograms of the empirical conditional
PDF's of a return $r_2$ once a previous return of modulus $|r_1|$ 
has been realized,
are very well reproduced by the analytical prediction based on
Eq. (\ref{solution_D0p5}) (Fig. \ref{fig_conditional}). 
The analytical expression for 
$\tilde p_{2T}^{(2)}$, and thus of $p_{2T}(r_2||r_1|)$, 
is here obtained on the basis of a particular form of $\tilde{g}$
\cite{sokolov_1}, whose parameters have been fixed by a preliminary fit 
of $g$ as illustrated in Fig. \ref{fig_scaling}.
In Fig. \ref{fig_conditional} we show the conditional PDF for a given 
$|r_1|$ at different values of $T$.

A further reason in favor of our solution is the fact that
the recipe in Eq. (\ref{solution_D0p5}) which constructs 
$\tilde p_{2T}^{(2)}$ through
replacement of single argument dependence of $\tilde{g}$ by a 
spherically symmetric dependence in $(k_1,k_2)$, can be regarded
as a rule of algebraic multiplication of $\tilde{g}(k_1)$
by $\tilde{g}(k_2)$. This type of multiplication, which is
straightforwardly generalized to more than two factors, is 
commutative and associative, has as neutral element $\tilde{g}(0)=1$,
and, most important, can be put at the 
basis of an extension of the central limit theorem for sums
of strongly correlated variables \cite{baldovin_2}. 
In this perspective
Eq. (\ref{solution_D0p5}) can be put in a form
\begin{equation}
\tilde p_{2T}^{(2)}(k_1,k_2)=
\tilde g\left(\sqrt{T}k_1\right)\otimes\tilde g\left(\sqrt{T}k_2\right),
\label{eq_gprod}
\end{equation}
where the symbol $\otimes$ represents the non-standard
multiplication \cite{baldovin_2}. 
The $\otimes$ multiplication reduces to the standard one for Gaussian
$g$'s \cite{baldovin_1,baldovin_2}, and in this case
Eq. (\ref{eq_gprod}) gives 
$\tilde p_{2T}^{(2)}(k_1,k_2)=
\tilde g\left(\sqrt{T}k_1\right)\;\tilde g\left(\sqrt{T}k_2\right)
$, consistently with Eq. (\ref{eq_mult}). 

\section{Non-stationary stochastic model of index evolution}
We already mentioned that the above solution in Eq. (\ref{solution_D0p5})
can be generalized to the case of more than two successive
intervals. Indeed, through our recipes we can construct a joint PDF
for many returns. If, for example, we consider
the CF corresponding to the joint PDF of $n$ successive daily returns,
$p_{n1}^{(n)}$,
this will be simply given by
\begin{eqnarray}
\tilde p_{n 1}^{(n)}(k_1,k_2,\ldots,k_n)&=& 
\tilde{g}(k_1)\otimes\tilde{g}(k_2)\otimes\cdots\otimes\tilde{g}(k_n)\label{eq_procD0p5a}\\
&=&\tilde g\left(\sqrt{k_1^2+k_2^2+\cdots+k_n^2}\right). 
\label{eq_procD0p5}
\end{eqnarray}
The joint PDF's
for various $n$ define a non-Markovian, self-similar stochastic 
process thanks to the algebraic properties of the $\otimes$ 
multiplication. It would be tempting to regard this process
as the one which directly generated the history of the index
to which $\tilde{g}$ pertains. Indeed, a nice property one can deduce
for it is that for such a process returns would be stationary.
So, the ensemble PDF for returns in the process could be directly
sampled by the sliding interval procedure yielding the empirical 
$p_T$.  Another nice property is that the process would be a martingale
\cite{feller_1}, i.e. the conditional expectation of the future return is always zero,
independent of the conditioning history. This is embodied in the construction
of the $\tilde p_{n1}^{(n)}$, which are even in the their dependence on each
of the $k_i$'s.
However, there are clear indications that this simple
scenario would be oversimplified, and that PDF's like
$p_{n1}^{(n)}$ cannot 
directly describe 
the postulated stochastic process 
potentially able to generate a whole ensemble
of alternative histories.

The reasons why such a process with stationary increments would
not be acceptable are two. In first place, for such a process
the autocorrelation function of the absolute value of daily returns
would not decrease with time, but rather be a constant, in disagreement with
the empirical observation of a power law decay ( Fig. \ref{fig_corr}a) 
\cite{cont_1}. Furthermore,
one has to consider that the construction of $p_{n1}^{(n)}$
assumes a strict simple-scaling form for $p_T$. We know that
this simple-scaling is well obeyed by this PDF only if we restrict
ourselves to consider its lowest moments, i.e. $\langle|r|^q\rangle_{p_T}$ with
$q \lesssim 3$, like we did in our argument leading to $D=1/2$.
In fact multiscaling-like effects are observed in the scaling of
higher moments \cite{baldovin_1,di_matteo_1}. 
These effects mean that the $q$-th moment of $p_T$ scales as
$T^{D(q)}$, with $D(q)<q/2$ for $q\gtrsim 3$, and 
would not be taken into account by the 
stochastic model.

A way out of the above difficulties is found if one considers that
the very assumption of stationary increments for
the process underlying index evolution is, a priori, not justified.
Stationarity is often assumed on the basis of
the fact that $p_T$ is stationary by construction. However, there is no
compelling reason to do this and to identify $p_T$ with the
PDF of the returns of the underlying process. The recent literature even reports 
indications
that the stochastic processes driving exchange rates
could be characterized by time-inhomogeneities in the returns
\cite{bassler_1}. 
Within our scheme it is easy to embody the possibility
of non-stationary returns in the ensemble generating process.
The key is found going back to our arguments leading to
the conclusion that $D=1/2$ for the empirical $p_T$. 
For the postulated non-stationary process
driving the index, the ensemble PDF for
returns like $r_{t,T}$ should be a function of both $t$ and $T$, which 
we indicate here by $P_{t,T}(r)$. Likewise, the joint PDF of the process
corresponding to $p_{2T}^{(2)}$ can be indicated by 
$P_{t,2T}^{(2)}(r_1,r_2)$. Let us now consider the equations
applying to these ensemble PDF's and corresponding to
Eqs. (3) and (4) for the empirical PDF's. With $t=0$
for simplicity, one gets:
\begin{eqnarray}
P_{0,T}(r_1)&=&\int d r_2 P_{0,2T}^{(2)}(r_1,r_2),\label{eq_r1}\\
P_{T,T}(r_2)&=&\int d r_1 P_{0,2T}^{(2)}(r_1,r_2).\label{eq_r2}
\end{eqnarray}
Eq. (\ref{prob_emp}) is simply rewritten as 
\begin{equation}
P_{0,2T}(r)= \int d r_1 d r_2\;P_{0,2T}^{(2)}(r_1,r_2)\;\delta(r-r_1-r_2).
\end{equation}
Suppose further that a simple-scaling form is valid for $P_{0,T}$,
i. e. 
\begin{equation}
P_{0,T}=\frac{1}{T^{D_e}}g_e\left(\frac{r}{T^{D_e}}\right), 
\label{eq_ens_scaling}
\end{equation}
with an ensemble scaling function $g_e$ and an
ensemble dimension $D_e$. Consider now the ensemble average
$\langle(r_1 +r_2)^2\rangle_{P_{0,2T}^{(2)}}$ in the light of 
Eqs. (\ref{eq_r1}-\ref{eq_ens_scaling}) and of 
$\langle r_1r_2\rangle_{P_{0,2T}^{(2)}}=0$. 
If we want to recast the
r.h.s. of Eq (\ref{eq_r2}) in the form of $P_{0,T'}$, we realize that we must put
$T' = a T$, with $a=(2^{2D_e}-1)^{1/2D_e}$, in order to be consistent with the scaling 
assumed for $P_{0,T}$. If $D_e \neq 1/2$,
there is an inhomogeneity in the process, measured by this rescaling $a$,
which reveals an asymmetry between the first and the second $T$-interval
considered. In other words, $a \neq 1 $ signals a
preferential direction of time: the evolution of the index in the second 
interval occurs with width rescaled with respect to that of the previous
one. This effect is consistent with causality and the rescaling
$T'=a T$ is analogous to the rescaling of the block size $N$
which one would have when constructing marginal PDF's 
for $M_1$ or $M_2$ in the magnetic RG. Of course,
in that case, there would not be causality, and the rescaling would
apply symmetrically to both PDF's as a consequence of the correlations. 

\begin{figure}[htbp]
\epsfxsize=8cm
\centerline{\epsfbox{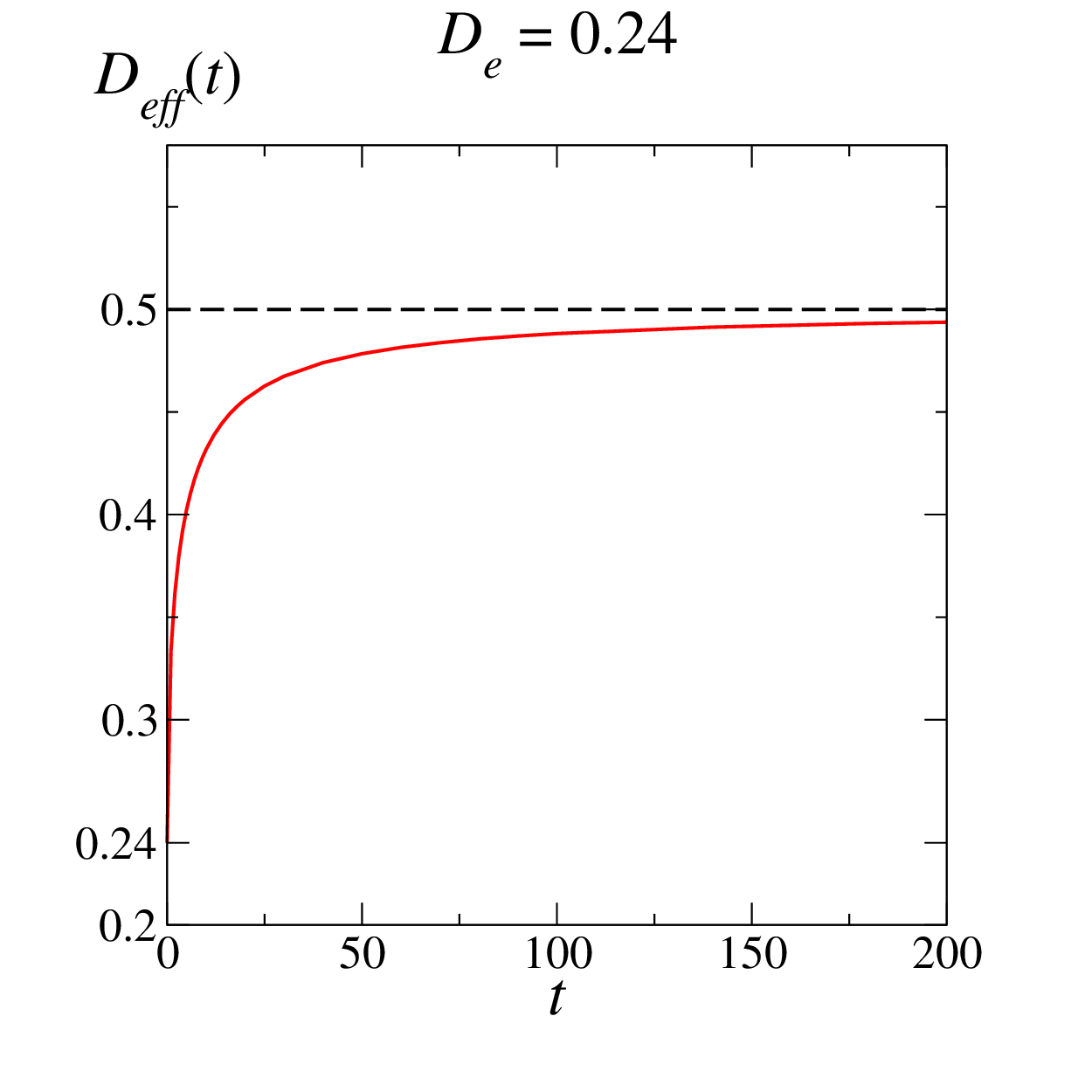}}
\caption{
  Effective dimension $D_{eff}$ for $1\lesssim T\lesssim 40$ deduced from
  Eq. (\ref{eq_inhom}). This $D_{eff}$ is obtained from the behavior of
  the moments of $P_{t,T}$ as a function of $T$ in this range.
}
\label{fig_Deff}
\end{figure}

This argument already shows that our formalism leaves room for
the construction of stochastic processes which are more general than
that the one introduced in Eqs. (\ref{eq_procD0p5a}-\ref{eq_procD0p5}). 
One has to follow steps similar to those
which led us to construct the solution for $\tilde{p}_{2T}^{(2)}$
in terms of $\tilde{g}$, taking into account the presence of $a \neq 1$.
For the CF of $P_{0,2T}^{(2)}$, for example, we get now
$\tilde{P}_{0,2T}^{(2)}(k_1,k_2)= 
\tilde g_e\left(T^{D_e}k_1\right)\otimes \tilde g_e\left((a T)^{D_e}k_2\right)$.
Similarly, for a sequence of $n$ daily returns, we can write:
\begin{equation}
\tilde P_{0,n1}^{(n)}(k_1,k_2,\ldots,k_n)=
\tilde g_e\left(a_1^{D_e}k_1\right)\otimes\tilde g_e\left(a_2^{D_e}k_2\right)
\otimes\cdots\otimes g_e\left(a_n^{D_e}k_n\right),
\label{eq_P}
\end{equation}
where $a_i\equiv\left[i^{2D_e}-(i-1)^{2D_e}\right]^{1/2D_e}$; $i=1,2,\dots n$.
These last CF's again fully characterize a stochastic process, which 
is now non-stationary. The process is consistent with the 
simple-scaling of $P_{0,T}$, but now $P_{t,T}$ satisfies a more general,
inhomogeneous form of scaling:
\begin{equation}
P_{t,T}(r)=\frac{1}{\sqrt{(t+T)^{2D_e}-t^{2D_e}}}\;\;
g_e\left(\frac{r}{\sqrt{(t+T)^{2D_e}-t^{2D_e}}}\right).
\label{eq_inhom}
\end{equation}
On the basis of this last equation, one can try to analyze
how the effective scaling dimension for $T$ of the order of the month
varies as a function of $t$. This is illustrated in Fig. \ref{fig_Deff}
for a value of $D_e=0.24$, which, as discussed below, is
directly relevant for the application to the DJI index.
One sees that, after a rather fast increase at short $t$,
this effective dimension $D_{eff}$ approaches the value $1/2$, which is 
the asymptotic limit. 
Thus, if one would sample with sliding interval method a return PDF $p_T$
along a sufficiently long single history of the process consistent with Eq. (\ref{eq_inhom}),
this PDF would show a scaling with $D \simeq 1/2$ for
$T$ of the order of the month. At the same time, the initial
deviation from $1/2$ of the effective dimension shown in 
Fig. \ref{fig_Deff} suggests that this $p_T$ could
manifest multiscaling-like features. 

The problem arises now of postulating some
concrete mechanism through which inhomogeneity 
can act in generating an index history. It is natural to assume that
the inhomogeneity crosses over to homogeneity when $t$  
exceeds some cut-off $t_c$. This cut-off time could be of the order
of the autocorrelation time of volatility, i.e. several hundreds 
of days. Of course, $t_c$ should be regarded as a statistical 
average of the duration of many random
intervals within which the process is described by a $P_{0,n1}^{(n)}$
corresponding to Eq. (\ref{eq_P}). At the junctions between these intervals,
which can be imagined to coincide with relevant external events 
influencing the market, one could assume that the progression of 
the coefficients $a_i$ is suddenly interrupted and restarted, either
from the beginning ($a_{i+1}=a_1$), or from a randomly chosen stage
($a_{i+1} = a_k$, with $k \neq 1$).

\section{Simulation of the model and results for the DJI index}

The knowledge of $P_{n1}^{(n)}$ allows to implement an
autoregressive strategy for the simulation of the process.
Autoregressive methods are used extensively in finance, e.g. for
the implementation of ARCH or GARCH processes \cite{engle_1,bollerslev_1}.
Suppose we consider $P_{n1}^{(n)}$, with, e.g., $n=100$.
If we give as input the first $99$ returns, the joint PDF can 
be used in order to define the conditional PDF of the
return in the hundredth day. Once this return is extracted
consistently, one can use the returns from the second to the 
hundredth day included in order to extract in a similar way the 
return of the $101$-th day, and so on. Without entering 
into the details of how this is practically implemented \cite{baldovin_1},
here we just review the results one can obtain.

By expressing $P_{n1}^{(n)}$ in terms of $\tilde{g}_e$, whose
expansion around $k=0$ is directly linked to that of 
the empirical $\tilde{g}$ \cite{baldovin_1}, we generate single histories
supposed to imitate the one at our disposal with over one century
of DJI index daily closures. In all cases we fix $t_0 = 500$ 
on average for the inhomogeneity updating and $n=100$ for 
the auto-regressive scheme. Once a single history is generated,
we act on it, by sliding interval sampling techniques,
exactly in the same way one does on the true historical series.
Upon varying $D_e$, which is the crucial parameter
in the simulation, we search for the most realistic behavior of the
empirical volatility autocorrelation function in the range from
few to about hundred days. Remarkably enough, for $D_e<1/2$ the obtained
autocorrelation functions behave as decaying power laws in this
range, and the exponent becomes very close to the empirical
one ($\beta \simeq 0.2$) for $D_e \simeq 0.24$, 
(see Fig. \ref{fig_corr}b and \cite{baldovin_1}). 
At the same time, we can try to optimize $D_e$ by requesting a 
realistic agreement between the multiscaling features of
the simulated $p_T$ and those of the empirical one. It
is remarkable that the optimal $D_e$ according to this second 
criterion is very close again to $0.24$ \cite{baldovin_1}. This is a
further indication that the model is coherent and catches the
essential statistical features of a long index history. 

\begin{figure}[htbp]
\epsfxsize=8cm
\centerline{\epsfbox{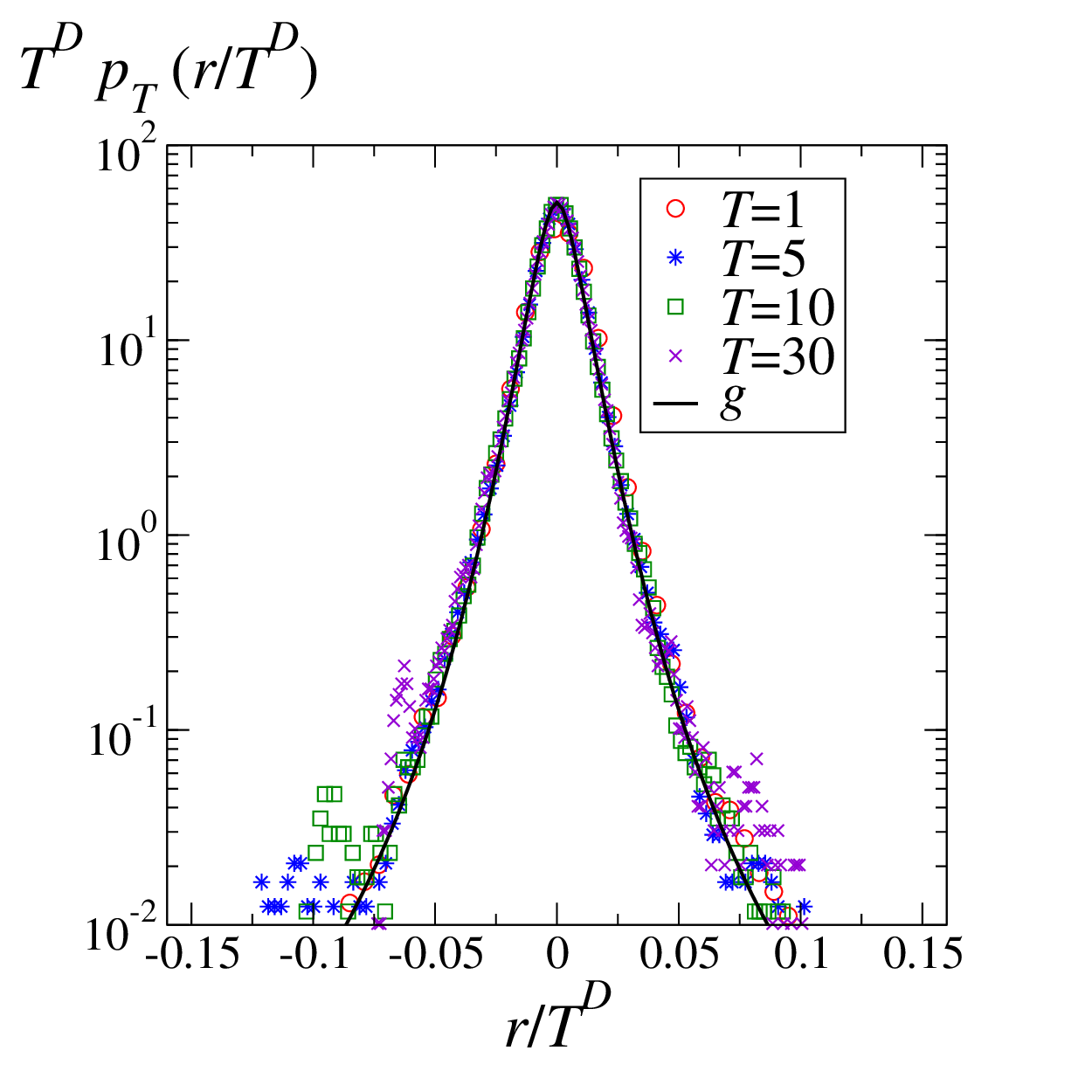}}
\caption{
  Scaling collapse of the $p_T$ sampled from a single simulated history.
}
\label{fig_scaling_sim}
\end{figure}

A very important test for the proposed model concerns the
scaling of the empirical $p_T$ itself. A crucial limitation of
simulation methods like ARCH and GARCH is that they generate 
histories for which it is not guaranteed that the sampled $p_T$
satisfies scaling. 
This is in fact regarded as a major open problem of these
approaches\cite{mantegna_1}. In Fig. \ref{fig_scaling_sim} we report the 
scaling collapse of $p_T$ as obtained by one of the histories
generated by our simulation. The collapse is clearly of the same
quality as that reported in Fig 1; moreover, the scaling function
and the exponent emerging from the collapse are very consistent with
those valid for the historical data. 

The general agreement between the stylized features of
the simulated histories and those of the the true history \cite{baldovin_1},
suggest that our model based on inhomogeneous scaling and market 
efficiency catches the robust features of the stochastic component
of index evolution.

\section{Conclusions}

The results reviewed in this report show the importance
of scaling in building up a model of stochastic 
index evolution in finance. In the construction of the model
this symmetry enters as a very crucial tool, in the sense that,
combined with other constraints, it leads to fix very plausible and
consistent rules of probabilistic evolution. 
In view of the success of the 
model, the skepticism on the practical relevance of scaling
mentioned in the introduction, should be attenuated. A crucial 
factor which helps in converting scaling into a powerful 
predictive tool here, is that we regard it in a 
perspective which has roots in the RG 
approach to criticality. This allows even to establish
novel paradigms, like the one represented by the inhomogeneous
scaling in Eq. (\ref{eq_inhom}). 
More generally, the strategy followed here 
tries to profit of the lessons learned from decades of work in complex
systems. As clearly stated in Ref. \cite{kadanoff_2}, the issue of 
financial market modeling should be addressed by first trying to
focus on the universal phenomenological features, and on the basic
symmetries, rather then privileging analytical tractability. 

We believe that the approach presented here could have more 
general applicability. There are many natural phenomena characterized
by anomalous scaling, for which part of the features
of the model discussed here, or of the arguments leading
to it, could reveal worth considering. These problems
belong in general to the fields of multidisciplinary applications of 
statistical mechanics.

\end{document}